\documentclass[12pt]{article}
\usepackage{graphicx}
\usepackage{subfigure}
\newcommand{\beq}{\begin{equation}}
\newcommand{\eeq}{\end{equation}}
\newcommand{\bea}{\begin{eqnarray}}
\newcommand{\eea}{\end{eqnarray}}

\newcommand{\ra}{\right\rangle}
\newcommand{\la}{\left\langle}

\newcommand{\overlrarrow}[1]{\vbox{\ialign{##\cr\cr
                  \leftrightarrowfill\crcr\noalign{\kern-1pt\nointerlineskip}
                  $\hfil\displaystyle{#1}\hfil$\crcr}}}

\begin{document}
\begin{titlepage}
\begin{flushleft}
       \hfill                      {\tt hep-th/0708.3706}\\
       \hfill                       FIT HE - 07-03 \\
       \hfill                       KYUSHU-HET ** \\
       \hfill                       Kagoshima HE - 07-2 \\
\end{flushleft}
\vspace*{3mm}
\begin{center}
{\bf\LARGE D3/D7 holographic Gauge theory \\
\vspace*{3mm}
and Chemical potential}

\vspace*{5mm}
\vspace*{12mm}
{\large Kazuo Ghoroku\footnote[2]{\tt gouroku@dontaku.fit.ac.jp},
Masafumi Ishihara\footnote[3]{\tt masafumi@higgs.phys.kyushu-u.ac.jp},
Akihiro Nakamura\footnote[4]{\tt nakamura@sci.kagoshima-u.ac.jp}
}\\
\vspace*{2mm}

\vspace*{2mm}

\vspace*{4mm}
{\large ${}^{\dagger}$Fukuoka Institute of Technology, Wajiro, 
Higashi-ku}\\
{\large Fukuoka 811-0295, Japan\\}
\vspace*{4mm}
{\large ${}^{\ddagger}$Department of Physics, Kyushu University, Hakozaki,
Higashi-ku}\\
{\large Fukuoka 812-8581, Japan\\
\vspace*{4mm}
{\large ${}^{\S}$Department of Physics, Kagoshima University, Korimoto 1-21-35,Kagoshima 890-0065, Japan\\}}

\vspace*{10mm}
\end{center}

\begin{abstract}
N=2 supersymmetric Yang-Mills theory with flavor hypermultiplets at finite temperature 
and in the dS${}_4$ 
are studied for finite quark number density ($n_b$) by a dual supergravity 
background with non-trivial dilaton and axion. The
quarks and its number density $n_b$ are introduced by embedding a probe D7 
brane. We find a critical value of the chemical potential at the limit
of $n_b=0$, and it coincides with the effective quark mass given
in each theory for $n_b=0$. 
At this point, a transition of the D7 embedding configurations occurs
between their two typical ones. The phase diagrams of this transition
are shown in the plane of chemical potential versus 
temperature and cosmological constant for YM theory at finite temperature 
and in dS${}_4$ respectively. In this phase transition, the order parameter 
is considered as $n_b$. 
This result seems to be reasonable since both theories are in the
quark deconfinement phase.

\end{abstract}
\end{titlepage}

\section{Introduction}

In the context of the holography \cite{juan,bigRev},
the dynamical properties of
flavor quarks have been studies by embedding D7 brane(s) as probe.
~\cite{KK,KMMW,KMMW2,Bab,ES,SS,NPR,GY,CNP}. 
Recently, the research in this direction has been extended
by introducing quark number density. Such researches have been performed 
at finite temperature for D3/D7 \cite{korea,Myers,KO} and D4/D8 
\cite{HT} models, and some kinds of phase transitions are reported. 

In D3/D7 model,
the high temperature gauge theory is represented by the AdS Schwartzchild
configuration, which describes chiral symmetric\footnote{
Here and hereafter we use the chiral symmetry for the state with 
massless quark or a hypermultiplet, and the breaking of this symmetry means
the spontaneous mass generation for the quark.} 
and deconfinement phase for
zero quark number ($n_b$). A horizon appears in the bulk configuration 
in this model, then two types of D7 embeddings are possible, the 
``Minkowski embedding'', in which the D7 brane is off the horizon, and 
``black hole embedding'' (BH), where the IR end point of the D7 brane touches on the 
horizon.  

For the case of non-zero $n_b$, however, the authors of \cite{Myers}
have pointed out that the Minkowski embedding is unphysical due to the
fact that the inner electric field remains 
finite at the IR end point of D7 brane, and it can not go nowhere
from that point without any appropriate source like a baryon. Therefore,
the physical embedding is restricted to the BH type, and they found
that all the quark mass range could be covered by the BH embedding.
This implies that the BH embedding is enough for finite $n_b$ case.

However, we show here that a finite region of chemical potential 
($0<\mu<\mu_0$) can not be covered by finite $n_b$ and BH embedding, and
this region is described rather by the physical Minkowski embedding with $n_b=0$. 
And the critical point $\mu_0(=\mu_0(T))$,
which depends on the temperature $T$, is shown to be equivalent to the effective
quark mass ($\tilde{m}_q(=\tilde{m}_q(T))$). 
Then $n_b$ appears when $\mu$ exceeds $\tilde{m}_q$.
This result implies that $n_b$ describes the 
number of quarks moving freely with the mass $\tilde{m}_q$ 
in the thermal YM medium. This interpretation 
is very reasonable since the chemical potential is
examined here for AdS Schwartzchild background which is dual to the gauge theory
in the high temperature deconfinement phase. 
Then it would be an open problem here how $n_b$ could be related to
the baryon number density of a physical baryon.
And the curve $\mu_0(T)$ in the
$\mu$-$T$ plane indicates the critical line which separates two phases,
Minkowski embedding with $n_b=0$ and BH embedding with $n_b\neq 0$. In passing this
critical curve, the chemical potential does not jump but the electric potential
and profile function of D7 embedding show a jump near the horizon. The order
parameter of this transition is interpreted as 
$n_b\propto \langle\Psi^{\dagger}\Psi\rangle$, where $\Psi$ denotes 
the quark field.

Similar situation to the finite temperature gauge theory is seen
also for the gauge theory in the de Sitter space time ($dS_4$) \cite{H,GIN}.
This theory is characterized by the
4d cosmological constant ($\lambda$), and the configuration of
the gravity dual has a horizon as in the high temperature theory. 
Then both the ``Minkowski'' and BH D7 brane embeddings are possible.
The dynamical role of $\lambda$ is similar to the temperature so far. In fact, 
we find for finite $\lambda$ that the chiral symmetry
is restored
and the quarks are deconfined.
Then it is useful to study the meaning of the ``baryon number'' $n_b$ also
in the $dS_4$ gauge theory. Actually, we find a similar critical curve
$\mu_0(\lambda)$ in the $\mu$-$\lambda$ plane and the same kind of phase
transition on this critical curve. We also show 
$\mu_0(\lambda)=\tilde{m}_q(\lambda)$,

where $\tilde{m}_q(\lambda)$ represents the effective mass of a quark
moving freely in dS${}_4$ space. 

\vspace{.3cm}
Thus throughout the gauge theory at finite temperature and in $dS_4$,
we find that the chemical potential remains finite as
$\mu_0$ in the limit of $n_b=0$ in a definite region of $T$ or $\lambda$. 
In other words,
there is an area of $n_b=0$ and finite $\mu(<\mu_0)$ in $\mu$-$T$ and
$\mu$-$\lambda$ planes.
In this area, the D7 brane is embedded by Minkowski type, but
the finite chemical potential plays no dynamical role since it disappears in the
brane action since the electric potential is given by a constant
$\mu$. However, when it exceeds a critical value  
$\mu_0$, $n_b(\propto \langle\Psi^{\dagger}\Psi\rangle)$ appears as a non-zero value
and the D7 profile is changed to BH type. This phenomenon could be
interpreted as a phase transition with the order parameter 
$\langle\Psi^{\dagger}\Psi\rangle$.
Our purpose is to give such phase diagrams and to discuss this
transition observed in the deconfinement phase.

\vspace{.2cm}
In section 2, we give the setting of the bulk background and the D7 brane
action and the equations of motion of embedding are given in section 3.
In the section 4, the chemical potential and its relation to the phase transition
is given, and phase diagrams are given in the section 5 and 6.
The summary is given in the final section.

\section{Background geometry}

We start from
10d IIB model retaining the dilaton
$\Phi$, axion $\chi$ and selfdual five form field strength $F_{(5)}$.
Under the Freund-Rubin
ansatz for $F_{(5)}$, 
$F_{\mu_1\cdots\mu_5}=-\sqrt{\Lambda}/2~\epsilon_{\mu_1\cdots\mu_5}$ 
\cite{KS2,LT}, and for the 10d metric as $M_5\times S^5$ or
$ds^2=g_{MN}dx^Mdx^N+g_{ij}dx^idx^j$, we find the solution.
The five dimensional $M_5$ part of the
solution is obtained by solving the following reduced 5d action,
\beq
 S={1\over 2\kappa^2}\int d^5x\sqrt{-g}\left(R+3\Lambda-
{1\over 2}(\partial \Phi)^2+{1\over 2}e^{2\Phi}(\partial \chi)^2
\right), \label{5d-action}
\eeq
which is written 
in the string frame and taking $\alpha'=g_s=1$. 

\vspace{.3cm}
\noindent{\bf Two deconfinement solutions}

The following solutions are obtained under the ansatz,
\beq
\chi=-e^{-\Phi}+\chi_0 \ ,
\label{super}
\eeq
which is necessary to obtain a supersymmetric solutions. However,
the supersymmetry is broken in the following two solutions due to
the cosmological constant $\lambda$ and the temperature $T$.

\vspace{.3cm}
\noindent{\bf (2) High temperature phase:  } First, we consider the high temperature
gauge theory given in \cite{GSUY}, and it is written as
$$ 
ds^2_{10}=G_{MN}dX^{M}dX^{N} ~~~~~~~~~~~~~~\hspace{6.5cm}
$$ 

\beq
=e^{\Phi/2}
\left\{
{r^2 \over R^2}\left(-f^2(r)dt^2+(dx^i)^2\right)+
{1\over f^2(r)}\frac{R^2}{r^2} dr^2+R^2 d\Omega_5^2 \right\} \ , 
\label{finite-T-sol}
\eeq 
\beq
e^\Phi= \left( 1+\frac{q}{r_T^4}\log({1\over 1-(r_T/r)^4}) \right) \ , 
\label{dilaton2}
\eeq
\beq
  f(r)=\sqrt{1-({r_T\over r})^4} , \label{tempe}
\eeq
where where $M,~N=0\sim 9$ and
$R=\sqrt{\Lambda}/2=(4 \pi N)^{1/4}$. 
The temperature $T$ is related to the parameter $r_T$ as $r_T=\pi R^2 T$,
and $q$ represent the VEV of gauge fields condensate~\cite{GY},
$q\propto \langle F_{\mu\nu}^2\rangle$.

\vspace{.3cm}
\noindent{\bf (2) dS${}_4$:  } The second solution is for
the dS${}_4$ solution \cite{GIN},
\beq
ds^2_{10}=e^{\Phi/2}
\left\{
{r^2 \over R^2}A^2\left(-dt^2+a(t)^2(dx^i)^2\right)+
\frac{R^2}{r^2} dr^2+R^2 d\Omega_5^2 \right\} \ , 
\label{finite-c-sol}
\eeq 
\beq
e^\Phi= 1+\frac{q}{r^4}{1-(r_0/r)^2/3\over (1-(r_0/r)^2)^3} \ , 
\label{dilaton}
\eeq
\beq
  A=1-({r_0\over r})^2, \quad a(t)=e^{2{r_0\over R^2} t}
\eeq
 The horizon is denoted by
$r_0$, which is related to the 4d cosmological constant $\lambda$  
as
\beq
  \lambda=4{r_0^2\over R^4}.
\eeq
In this configuration, the four dimensional boundary represents the 
$\cal{N}$=4 SYM theory in the de Sitter space or in the inflational
universe characterized by the 4d cosmological constant $\lambda$.

\vspace{.3cm}
\section
{\bf D7 brane action and equation of motion:}~
The embedding of the brane is performed by solving the equations of motion
for the fields on the brane. 
\vspace{.3cm}
The brane action for the D7-probe is given as
$$ 
S_{\rm D7}= -\tau_7 \int d^8\xi \left(e^{-\Phi}
    \sqrt{-\det\left({\cal G}_{ab}+2\pi\alpha' F_{ab}\right)}
      -{1\over 8!}\epsilon^{i_1\cdots i_8}A_{i_1\cdots i_8}\right)
$$
\beq
   +\frac{(2\pi\alpha')^2}{2} \tau_7\int P[C^{(4)}] \wedge F \wedge F\ ,
\label{D7-action}
\eeq
where $F_{ab}=\partial_aA_b-\partial_bA_a$.
${\cal G}_{ab}= \partial_{\xi^a} X^M\partial_{\xi^b} X^N G_{MN}~(a,~b=0\sim 7$,
and $M,~N=0\sim 9$)
and $\tau_7=[(2\pi)^7g_s~\alpha'~^4]^{-1}$ represent the induced metric and
the tension of D7 brane respectively.
And $P[C^{(4)}]$ denotes the pullback of a bulk four form potential,
\beq
C^{(4)} = 
\left(\frac{r^4}{R^4} d x^0\wedge d x^1\wedge
d x^2 \wedge d x^3 \right)\ .
\label{c4}
\eeq
The eight form potential $A_{i_1\cdots i_8}$,
which is the Hodge dual to the axion, couples to the
D7 brane minimally. In terms of the Hodge dual field strength,
$F_{(9)}=dA_{(8)}$ \cite{GGP}, the potential $A_{(8)}$ is obtained. 
In this brane action, there are two
scalar fields, which are denoted here as $X^8$ and $X^9$,
corresponding to the two coordinates transverse to the brane.
Since the brane action is rotational symmetric in the plane of these two
coordinate, the embedding is given by the solution of one of the scalar fields,
and it is denoted by $w(=X^8)$ given below.

\vspace{.3cm}
In addition, in the present case, we must solve for the Wick rotated
time-component ($A_0$)
of the $U(1)$ vector fields on the brane in order to introduce
the chemical potential ($\mu$) of the quark.

\vspace{.3cm}
\noindent{\bf For finite temperature :  }

\vspace{.3cm}
In the case of finite temperature, the D7 brane is 
embedded as follows \cite{GSUY}. The six dimensional part of the above metric
is rewritten as,
\beq
 {1\over f^2(r)}\frac{R^2}{r^2} dr^2+R^2 d\Omega_5^2
 =\frac{R^2}{U^2}\left(d\rho^2+\rho^2d\Omega_3^2+(dX^8)^2+(dX^9)^2
\right)\ ,
\eeq
\beq
  U(r)=\exp\left( \int{dr\over r\sqrt{1-(r_T/r)^4}} \right)
     =r\sqrt{{1+f(r)\over 2}}.
\eeq
Here $U$ is normalized as $U=r$ for $r_T=0$, and $U^2=\rho^2+(X^8)^2+(X^9)^2$.
Then we obtain the induced metric for D7 brane,
$$ 
ds^2_8=e^{\Phi/2}
\left\{
{r^2 \over R^2}\left(-f^2(r)dt^2+(dx^i)^2\right)+\right.
\hspace{3cm}
$$
\beq
\left.\frac{R^2}{U^2}\left((1+(\partial_{\rho}w^8)^2
+(\partial_{\rho}w^9)^2)d\rho^2+\rho^2d\Omega_3^2\right)
 \right\} \ , 
\label{D7-metric}
\eeq
where $w^8(\rho)$ and $w^9(\rho)$ are the scalars which determine
the position of D7 brane. They are solved under the ansatz that
they depend on only $\rho$. Further we can set $w^9=0$ and $w^8=w(\rho)$
without loss of generality due to the rotational invariance in
$X^8-X^9$ plane. 
By substituting the above metric in the DBI action and retaining $A_0(\rho)$, 
we obtain
\beq
S_{\rm D7}^{T} =-\tau_7~\int d^8\xi  \sqrt{\epsilon_3}\rho^3
\left(\left({r\over U}\right)^4 f
   e^{\Phi}\sqrt{ 1 + (w')^2-{U^2\tilde{F}_{\rho t, T}^2\over f^2r^2e^{\Phi}}}-
           {q\over U^4} \right)
\ ,
\label{D7-action-3}
\eeq
where $\tilde{F}_{\rho t}=2\pi\alpha'{F}_{\rho t}$, 
${F}_{\rho t}=\partial_{\rho}A_0$ and 
the last term in the parenthesis denotes
the eight form part.

The equations of motion are solved as follows. 
First, we solve the equation of motion for $A_0$ by introducing the quark
number density $n_b$ as an integral constant as follows,
\beq
{\rho^3(r/U)^2\tilde{F}_{\rho t,~T}\over
   f\sqrt{ 1 + (w')^2-{\tilde{F}_{\rho t,~T}^2\over f^2(r/U)^2e^{\Phi}}}}=n_b
\eeq
 From this, we obtain the following form of $U(1)$ electric field,
\beq
\tilde{F}_{\rho t, T}=n_b f{r\over U}e^{\Phi/2}
   \sqrt{ {1 + (w')^2\over {n_b^2+\rho^6 (r/U)^6e^{\Phi}}}}, \label{field-T}
\eeq

Here we see the property of $\tilde{F}_{\rho t, T}$ near the horizon
$r=r_T$. At this limit, we have
\beq
\tilde{F}_{\rho t, T}\sim n_b(1-{r_T\over r})
\eeq
for both $q=0$ and $q\neq 0$. Then $\tilde{F}_{\rho t, T}=0$ at the horizon,
while it is finite at $\rho=0$ for $n_b\neq 0$. So the only the BH embedding is
physical for $n_b\neq 0$, and the Minkowski embeddings are considered only for
$n_b=0$. 


\vspace{.3cm}
Finally, then, the equation of motion for $w$ is given by 
\bea
&&{{w''}\over{1+w'~^2}}+3{{w'}\over{\rho}}-{{w-\rho w'}\over{U^2f}}
\left[2(1-f)^2-{{4q}\over{r^4}}e^{-\Phi}\right]  \cr
\smallskip 
&&-{w{(1+w'~^2)}^{1/2}\over U^2f}
 {{4q}\over{r^4}}e^{-\Phi}(1-G_T)^{1/2}  \cr
\smallskip
&&-{{G_T}}
\left[{3w'\over{\rho}}-{{(w'\rho-w)}\over{U^2 f}}\left( 3(1-f)f
+{{2q}\over{r^4}}e^{-\Phi}\right)\right] 
=0\ ,
  \label{Teq}
\eea
where $G_T$ is defined as
\beq
G_T={{n_b^2}\over{n_b^2+\rho^6(r/U)^6 e^\Phi}}\ .
 \label{GTeq}
\eeq
Of course, this equation reduces to the one given in \cite{GSUY}
when $n_b\to 0$.

\vspace{.7cm}
\noindent{\bf For $dS_4$:  }

\vspace{.3cm}
In this case,
the extra six dimensional part of the above metric (\ref{finite-c-sol})
is rewritten as,
\beq
 \frac{R^2}{r^2} dr^2+R^2 d\Omega_5^2
 =\frac{R^2}{r^2}\left(d\rho^2+\rho^2d\Omega_3^2+(dX^8)^2+(dX^9)^2
\right)\ ,
\eeq
where $r^2=\rho^2+(X^8)^2+(X^9)^2$.
And we obtain the induced metric for D7 brane,
$$ 
ds^2_8=e^{\Phi/2}
\left\{
{r^2 \over R^2}A^2\left(-dt^2+a(t)^2(dx^i)^2\right)+\right.
\hspace{3cm}
$$
\beq
\left.\frac{R^2}{r^2}\left((1+(\partial_{\rho}w)^2)d\rho^2+\rho^2d\Omega_3^2\right)
 \right\} \ , 
\label{D7-metric-2}
\eeq
where we set as $X^9=0$ and $X^8=w(\rho)$ as mentioned above.
Then 
retaining the profile function $w(\rho)$ and $A_0(\rho)$, 
we arrive at the following D7 brane
action,
\beq
S_{\rm D7} =-\tau_7~\int d^8\xi  \sqrt{\epsilon_3}\rho^3 a(t)^3
\left(A^4
   e^{\Phi}\sqrt{ 1 + (w')^2-{\tilde{F}_{\rho t}^2\over A^2e^{\Phi}}}-C_8 \right)
\ ,
\label{D7-action-2}
\eeq
where $\tilde{F}_{\rho t}=2\pi\alpha'{F}_{\rho t}$, 
${F}_{\rho t}=\partial_{\rho}A_0$ and the eight form part
is given as
\beq
 C_8(r)=\int^rdr'~A^4(r')\partial_{r'}\left({\rm exp}({\Phi(r')})\right)
={q\over r^4} . \label{A8}
\eeq
The equations of motion are solved as above. 
First, we solve the equation of motion for $A_0$ by introducing the quark
number density $n_b$ as an integral constant, 
\beq
{\rho^3A^2\tilde{F}_{\rho t}\over
   \sqrt{ 1 + (w')^2-{\tilde{F}_{\rho t}^2\over A^2e^{\Phi}}}}=n_b
\eeq
 From this, we obtain the following form of $U(1)$ electric field,
\beq
\tilde{F}_{\rho t}=n_bAe^{\Phi/2}
   \sqrt{ {1 + (w')^2\over {n_b^2+\rho^6A^6e^{\Phi}}}}, \label{field}
\eeq
Then we get the equation of $w$ by substituting this into the one obtained from
D7 action by the variational principle. The final form is written as,
\bea
&&{1\over{\sqrt{1+w'~^2}}}\Biggl\{{{w''}\over{1+w'~^2}}+w'\left[{3\over{\rho}}
+(\Phi+4\log A)'-{G} 
\left({3\over{\rho}}+(3\log A+{\Phi\over 2})'\right)
\right] 
\Biggr\}  \cr
\smallskip
&&+{w\over{\rho+w w'}}\Biggl\{
   \sqrt{1-G}\Phi'-\sqrt{1+w'~^2}\Biggl[(\Phi+4\log A)'
-{G}({\Phi\over 2}+3\log A)'
\Biggr]\Biggr\}
\cr 
&=&0\ ,   \label{qeq}
\eea
where prime denotes the derivative with respect to $\rho$,  
and $G$ is defined as
\beq
G={{n_b^2}\over{n_b^2+\rho^6 A^6 e^\Phi}}\, .\label{dSG}
\eeq
When we take the limit $n_b\to 0$ at non-zero $\rho$,
the above equation (\ref{qeq}) reduces to the one given in \cite{GIN}. 
By giving $n_b$ as a parameter, 
we can solve this equation and find the profile of the embedded D7 brane. Then 
we find $\tilde{F}_{\rho t}$ simultaneously through (\ref{field}).

\vspace{.3cm} 
\section{Chemical potential and phase transition}
 From the equations given above, we firstly solve for $w$  
with the boundary condition, $w(\infty)=2\pi\alpha' m_q$, where
$m_q$ denotes the current quark mass and we set as $2\pi\alpha'=1$ hereafter.
Then we can read from the asymptotic form of this solution the chiral condensate 
$\la\bar{\Psi}\Psi\ra$, 
where $\Psi$ denotes the quark field. Then we solve for 
the gauge potential $A_0(\rho)$ by using this solved $w$.

In this case, we can assume
its asymptotic form at large $\rho$ as in the case of $w$.
In the context of AdS/CFT, it would be written in terms of
the chemical potential $\mu$ and the quark (or baryon) number density $n_b$ as,
\beq
A_0(\rho)= \mu-{n_b\over 2\rho^2}+\cdots \, ,\label{a0-eq}
\eeq 
We notice 
\beq
n_b\propto \langle \Psi^{\dagger}\Psi\rangle
   =\langle \bar{\Psi}\gamma^0\Psi\rangle\, .
\eeq
This would represent the baryon number density when the theory is in the
confinement phase, but we call this as quark number here since the present
model describes the deconfinement phase. 

In the present case, we firstly give $n_b$, then solve for $w(\rho,~n_b)$. And
by using this solution $w$, we obtain the chemical potential $\mu=A_0(\infty)$
from the following formula
\beq
 \tilde{\mu}=\int_{\rho_m}^{\infty}d\rho F_{\rho t}=A_0(\infty)-A_0(\rho_m)
        =\mu-A_0(\rho_m)
\, , \label{mu}
\eeq
where $\rho_m$ is the minimum value of $\rho$ and
we remind $\tilde{F}_{\rho t}=2\pi\alpha' {F}_{\rho t}$. 

In order to make
clear our viewpoint, we firstly restrict the analysis to the case 
of large $m_q$ where $w(\rho,~n_b=0)$ is the Minkowski embedding solution.
For $n_b=0$, $\tilde{F}_{\rho t}=0$, then the Minkowski embedding is
physical. This does not mean $\mu=0$, but it implies $A_0=\mu$ where
$\mu$ is conatant. In this case, $\tilde{\mu}=0$ and $A_0(\rho_m)=A_0(0)=\mu$.
So we can consider two types of solutions for $A_0$ depending on the solution
of $w$, (A) Minkowski type with $n_b=0$ and 
(B) black hole type with $n_b\neq 0$, in the case of large $m_q$ as given below.

\vspace{.5cm}
For the BH embedding solutions
with $n_b\neq 0$, the infrared boundary value $A_0(\rho_m)$
is estimated by using Eqs. (\ref{field-T}) and (\ref{field}).
For enough small $n_b$, $\rho_m$ is very small and the solution $w$ 
very rapidly increases from the point $\rho=\rho_m$. It soon
arrives at the value near the maximum value
$w(\infty)$ at about $\rho=\rho_m+\epsilon$, $\epsilon \ll 1$. In other 
words, $w$ is approximated by a step function then $w'$ is done by the $\delta$
function. As a result for $1\gg n_b \gg \rho_m^3$, we obtain
\beq
 \int_{\rho_m}^{\infty}d\rho ~F_{\rho t}\sim 
 \int_{\rho_m}^{\rho_m+\epsilon}d\rho ~A e^{\Phi/2}w'=
 \int_{r_0}^{w(\rho_m+\epsilon)}dr ~A e^{\Phi/2}
\, , \label{qmass-lam}
\eeq
for $dS_4$ with $q=0$, and 
\beq
 \int_{\rho_m}^{\infty}d\rho ~F_{\rho t, T}\sim 
 \int_{\rho_m}^{\rho_m+\epsilon}d\rho ~f{r\over U}e^{\Phi/2}w'=
 \int_{r_T}^{w(\rho_m+\epsilon)}dr ~e^{\Phi/2}
\, , \label{qmass-T}
\eeq
for finite temperature for any $q$.
Here we notice the dilaton denoted by $\Phi$ in the above two equations are
different from each other. Their explicit forms are given in the section two.
For the case of $dS_4$, however, the BH embedding is obtained only for $q=0$
since $F_{\rho t}$ diverges at $\rho_m$ for $q\neq 0$.
Meanwhile such a situation can not be seen for the finite temperature phase,
so we can consider any value of $q$ in this case. 
On these points, we discuss again in the below.

In the limit of $\rho_m=0$,
the most right hand side of (\ref{qmass-lam}) and (\ref{qmass-T})
are approximated as
\beq
\int_{r_0}^{w(0)}dr ~A e^{\Phi/2}=\tilde{m}_q\, ,
\quad \int_{r_T}^{w(0)}dr ~e^{\Phi/2}=\tilde{m}_{q, ~T}\, .\label{qmass}
\eeq
They are 
equivalent to the effective quark mass given in \cite{GIN} and
\cite{GSUY} defined from the Wilson line
in $dS_{4}$ and at finite temperature respectively for the Minkowski embedding
case with $F_{\rho t}=0$.\footnote {
The authors of \cite{Myers} discuss on this point related to
the string energy from a similar
viewpoint.
}
In both cases, the quarks are deconfined and moves
in the corresponding gauge field medium with this effective mass. 
On the other hand, the left hand side of (\ref{qmass-lam}) and (\ref{qmass-T})
, by its definition, is equivalent to $\mu-A_0(\rho_m)$ at $\rho_m\to 0$.
If we consider the chemical potential should be the effective quark mass
in the limit $n_b\to 0$, then
we should take as
\beq
 A_0(\rho_m)=0\, , 
\eeq
and we find in the limit of $n_b\to 0$
\beq
\tilde{\mu}(n_b\to 0)=\mu(n_b\to 0)\equiv \mu_0
=\tilde{m}_{q}\, ~~{\rm or} ~~\tilde{m}_{q, ~T}.
\label{eff-mass}
\eeq 
We must notice here that this analysis is performed for $n_b\neq 0$
even if the limit of $n_b\to 0$ is taken, and the embedding profile is 
assumed to be the BH type. 

\vspace{.5cm}
The important point to be noticed here is
that the value of $\mu_0$ is finite. Then there is a region,
$0<\mu<\mu_0$ and $n_b=0$, for finite
$\lambda$ or $T$ in both cases. In these regions, 
the Minkowski embedding is allowed, then we can consider as mentioned above
the following phase 
$$ {\rm Phase~~ (A)}~~n_b=0:~~
\left\{\rho_m=0\,, ~~A_0(0)=\mu\right\} $$
$$~~A_0=\mu={\rm const.}\, .$$
On the other hand for $\mu_0<\mu$, the following BH embedding phase (Phase (B)) is 
realized,
$$ {\rm Phase~~ (B)}~~n_b>0:~~
\left\{\rho_m>0\,, ~~A_0(\rho_m)=0\right\} $$
$$~~A_0(\rho)=\mu-{n_b\over 2\rho^2}+\cdots ~~{\rm at ~large}~\rho\, .$$
And the critical curves separating (A) and (B) are given by $\tilde{m}_q(\lambda)$ and
$\tilde{m}_{q,~T}(T)$ for each model. 

\vspace{.5cm}
For this definition of two phases, we show the phase diagrams in the next section
including numerical analyses. We should notice here that,
in this phase transition, from pahse (B) to (A), $A_0$ jumps from
$A_0(\rho_m\neq 0)=0$ to $A_0(0)=\mu_0$ at the limit of $\rho_m=0$. 
Simultaneously, the embedding profile $w$ changes from the BH
to the Minkowski type. Actually from the left
equations of the above approximate formula (\ref{qmass-lam}) and (\ref{qmass-T}),
we find $A_0'\sim w'$, then we can understand
the jump of $A_0$ corresponds to the jump of $w$ near
the horizon.

\section{Phase transition at high temperature phase}

\vspace{.3cm}
Here we firstly solve the embedding equations by introducing the chemical 
potential and noticing that any
Minkowski embedding solution is unphysical for $n_b\neq 0$.
As mentioned above, the Minkowski embedding is considered only for 
$n_b=0$. 
In this case, $\tilde{F}_{\rho t}=0$ at any point, then we find physical
Minkowski embeddings but there is no dynamical role of chemical potential
since it vanishes completely from the action.

\vspace{.3cm}

As stated above, two types of embeddings, (A) Minkowski and (B) BH embeddings, are
possible, and they are discriminated by the form of $A_0$ and $w$. And the critical
curve separating the two regions is given by $\mu_0=\tilde{m}_q$. Before giving
this critical curve, we discuss another transition which has been observed in
\cite{korea,Myers}.

\vspace{.3cm}
\noindent{\bf (i) Phase transition within (B) at small $\mu$:  } 

In the high temperature model, the topology changing phase transition
is observed for $q=0$ and $n_b=0$. This transition is characterized by the
jump of $w(\rho_m)$, where $\rho_m$ represents the minimum point of $\rho$.
In this sense, this transition is the first order since the D7 energy jumps
at this point due to the different configuration of $w$.
This kind of transition is also observed for finite $n_b$ between the same 
BH embedding. So this is not the topology changing phase transition in the
case of $n_b\neq 0$.

\begin{figure}[htbp]
\vspace{.3cm}
\begin{center}
  \includegraphics[width=10.5cm]{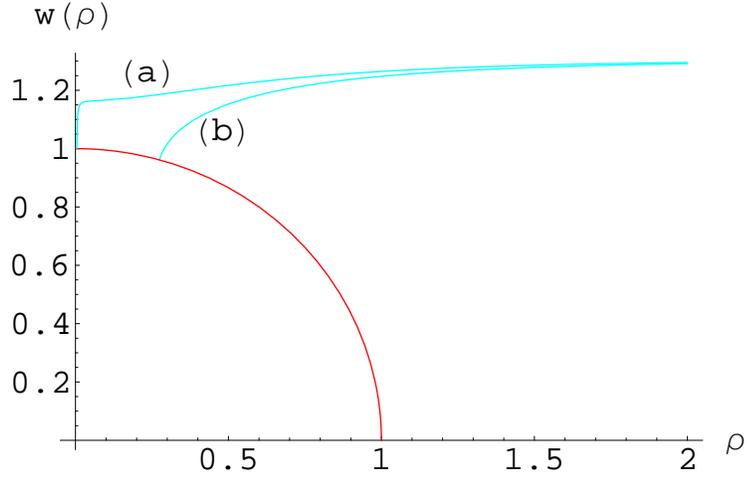}
\caption{Two embedding solutions for $q=0$, $m_q=1.309, n_b=10^{-5}$ at
the critical temperature $T_1=0.45$.
 \label{wq0fig}}
\end{center}
\end{figure}

{At the critical temperature $T_1$, we find
three different solutions at very small $n_b$
for the same quark mass, $m_q=w(\infty)$. They are shown
in the Fig. \ref{wq0fig}.} 
This is firstly observed in 
\cite{korea}. In \cite{korea}, $w$ is assigned
as the radial coordinate in $X^8$-$X^9$ plane, but this is not important.
We set as $w=X^8$, and we obtain almost the
same result with \cite{korea}.

\begin{figure}[htbp]
\vspace{.3cm}
\begin{center}
  \includegraphics[width=10.5cm]{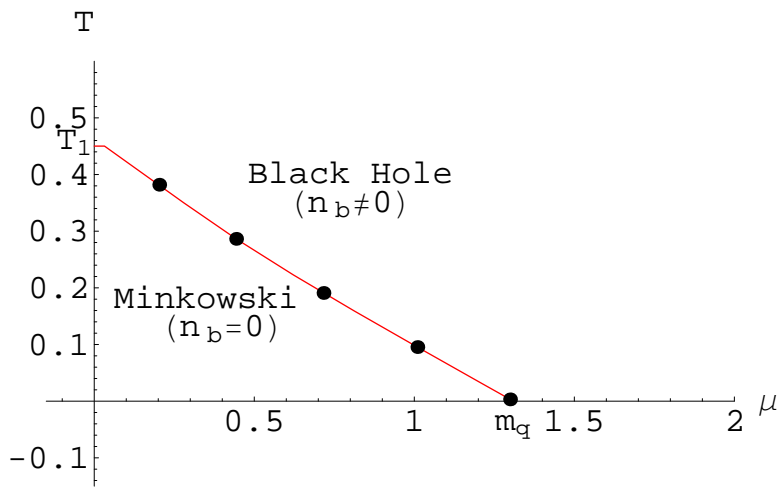}
  \caption{Phase diagram in T-$\mu$ plane for R=1,$m_q=1.30916$.
  The dots represent the 
  effective quark mass $\tilde{m}_q$ given by the last equation
  (\ref{qmass}). 
 \label{chemical1}}
\end{center}
\end{figure}

The solutions are separated to two
categories
whether the limit value
$\mu |_{n_b=0}(\equiv \mu_0)$ is finite (sol. (a)) or zero (sol. (b)). 
We notice here that
the two solutions for the case of solution (b) are overlapped in the figure
due to too small $n_b$.
Then, this gives a gap of 
$\mu_0(T)$ at this temperature $T_1$ 
as seen in the phase diagram in $\mu$-$T$ plane given
in the Fig. \ref{chemical1}.
The solution (a) provides finite $\mu_0$, and we observe 
$${\partial\mu\over\partial n_b} <0 ,$$
for this solution.
This fact implies that this solution is unstable as pointed out in \cite{Myers}.
And this is seen for the very small $\mu$ region of
$0<\mu<\mu_1$, where $\mu_1$ depends on $m_q$ and
is very small. 
As shown below, in the $dS_4$ model, 
there is no such a transition point at finite
$n_b$ so the gap is not seen. These points are assured in the phase diagrams
given below.

\begin{figure}[htbp]
\vspace{.3cm}
\begin{center}
  \includegraphics[width=10.5cm]{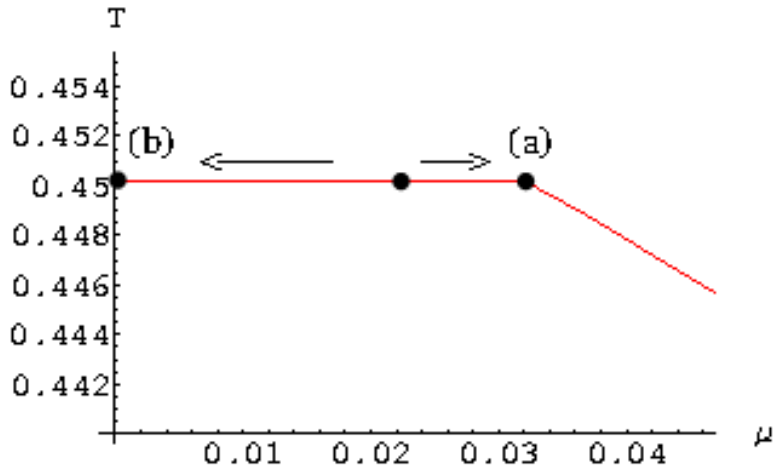}
  \caption{The extended small $\mu$ region of the Fig.
  \ref{chemical1} is shown. At first,
the two BH embeddings appear on the horizontal line at 
  $(T,\mu,n_b)=(0.450158,0.0223708,0.00699)$ in a degenerated form. 
Then for decreasing $n_b$,
  they run on the line in the opposite direction shown by the arrows and
  approache to the point (a) and (b), which corresponds to the two BH configurations
  shown in the Fig.\ref{wq0fig}.
 \label{chemical2}}
\end{center}
\end{figure}

Finally, we comment on the relation between the transition discussed here in 
terms of the Fig.\ref{wq0fig} and the one observed in the case of $n_b=0$. 
The solution (a) in the Fig.\ref{wq0fig} transits to the
Minkowski solution and the solution
(b) remains as BH solution in the limit of $n_b\to 0$. 
This situation of the two BH configurations is schematically depicted 
in the Fig.\ref{chemical2}. 
In this sense, the transition considered here
can be regarded as the limitting case of the
topology changing phase transition which is seen at $n_b=0$. 

\begin{figure}[htbp]
\vspace{.3cm}
\begin{center}
  \includegraphics[width=6.5cm]{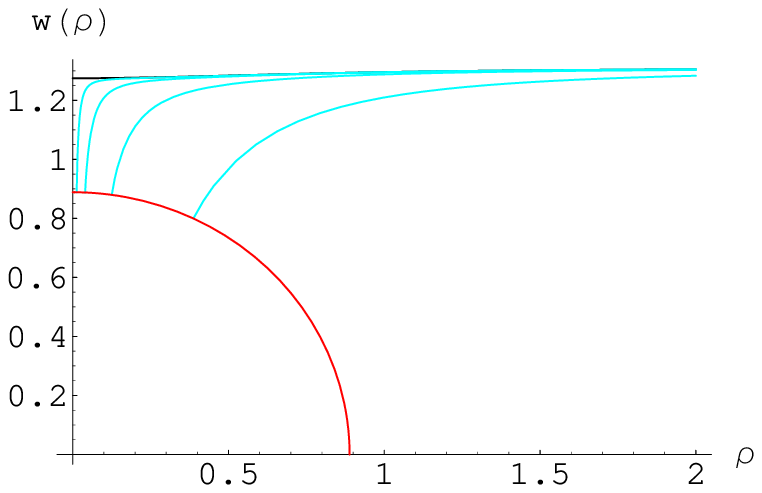}
  \includegraphics[width=6.5cm]{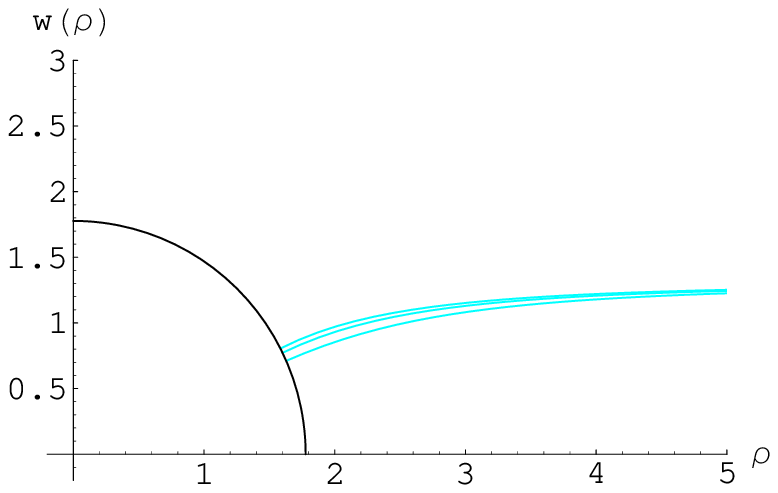}
 \caption{Embedding solutions near the transition point 
for $q=0$, $m_q=1.30916$. The left are for $T=0.4$, and 
$n_b=$0, 0.000187, 0.00184, 0.0181, 0.217 from the above.
The right is for $T=0.8$, and $n_b=$0, 5, 10.
 \label{chemT-q0}}
\end{center}
\end{figure}

\vspace{.3cm}
\noindent{\bf (ii) Phase transition from (B) to (A):  } 

We notice that $\mu_0(T)$ is zero for $T>T_1$ and it becomes finite 
in the low temperature side $T<T_1$. (See Fig. \ref{chemical1}.)
The situation is here considered from the viewpoint of grand canonical ensemble,
where the chemical potential $\mu$ is regarded as a parameter of the theory.
In this sense,
in the region of $0\leq \mu < \mu_0$, 
the solutions (A) (Minkowski embedding) with $n_b=0$ are found. 
When $\mu$ exceeds $\mu_0$,
$\mu$ becomes active and could generate a finite 
$n_b(\propto \langle\Psi^{\dagger}\Psi\rangle)$ to form the solution (B) 
(BH embedding). Then, for
$\mu_0<\mu$, the brane profile is changed to the BH embedding solution
when its D7 energy density becomes smaller than the one of the Minkowski embedding
with the same $\mu$ but with $n_b=0$. We show in the below
that this phenomenon really occurs at $\mu=\mu_0(T)$.

\vspace{.3cm}
This implies a topology change of the D7 embedded configuration in the bulk 10 
dimension, and, on the other hand, it is observed as the generation of $\langle\Psi^{\dagger}\Psi\rangle$
in the 4d gauge theory. 
The critical point $\mu_0(T)$ is determined for fixed 
$m_q=w(\infty)$ and $T$ 
by taking the limit of $n_b=0$ from the side of $n_b\neq 0$ (in the phase (B)).

\vspace{.3cm}
The typical solutions of $w(\rho)$ near this transition point for $(T,m_q)=(0.4,1.3)$
are shown for various $\mu$ in the left of the Fig.~\ref{chemT-q0}, where
one Minkowski solution with $n_b=0$
and four BH solutions with finite
values $n_b$ are shown. For the latter BH solutions,
we obtain the corresponding
values of $\mu(>\mu_0)$, where $\mu_0$ at this temperature ($T=0.4$)
is read from the Fig.~\ref{chemical1} as about $\mu_o=0.14$.
We notice that we have also the Minkowski solution for those value of $\mu$
with $n_b=0$.
Therefore, we must compare
the D7 energy of each BH embedding solution and the corresponding Minkowski solution
at the same value of $\mu$ in order to see which solution has lower energy.
The D7 energy $E_{\rm D7}^T$ is defined as 
\beq
S_{\rm D7}^T =-\tau_7~\int d^7\xi  ~\epsilon_3^{~1/2}E^T 
\ ,
\label{D7-energyT}
\eeq
then for the Minkowski embedding of $q=0$ we obtain
\beq
E_{M}^{T} =\int_0^{\infty} d \rho ~\rho^3
\left({r\over U}\right)^4 f
   e^{\Phi}\sqrt{ 1 + (w')^2}
\ ,
\label{D7-action-M}
\eeq
and for BH solution

\beq
E^{T}_{BH} =\int_{\rho_m}^{\infty} d \rho ~\rho^3
\left({r\over U}\right)^4 f
   e^{\Phi}\sqrt{ (1 + (w')^2)(1-G_T)}
\ ,
\label{D7-action-BH}
\eeq
where $G_T$ is given in (\ref{GTeq}). Each energy diverges, so we calculate
the difference of $E_{M}^{T}$ and $E^{T}_{BH}$, 
$$\Delta E^T=E_{M}^{T}-E^{T}_{BH}$$
to cancel the divergence coming from large $\rho$ integration. 
At large $\rho$, the solutions of
Mincowski and BH embeddings approaches to the same function, then this 
method of the regularization works well. And we could see 
$E_{M}^{T}>E^{T}_{BH}$ for any case, then the transition occurs at $\mu_0$
between the Minkowski solution and the one of the BH with the smallest $n_b$.
In other words, the
curve $\mu_0(T)$ obtained
in this way gives the critical curve which separates two phases (A) and (B),
which are defined above.

\vspace{.3cm}
In order to assure the statement, $\mu_0(T)=0$ for $T>T_1$, given above,
we also studied at $T=0.8$ where all the solutions are BH type even if 
$n_b=0$. The solutions are shown in the right hand side of Fig.~\ref{chemT-q0}.
In this case, we could see that $E^{T}_{BH}$ decreases with increasing $n_b$.
So no Minkowski embedding appears as a stable state. 

\vspace{.8cm}
The resultant curve of $\mu_0(T)$ is shown in the Fig.~\ref{chemical1}. 
Near $T=T_1\sim 0.45$, a small flat region is seen. This point is the transition
point shown in (i) above. The value of $\mu$ jumps from 0 to a small but finite 
$\mu_1$. Then, for $\mu_1<\mu$,
$\mu_0(T)$ increases monotonically with decreasing $T$, and 
arrives at $m_q=w(\infty)$ in the limit of $T=0$. However, this diagram
has its meaning only for $T>T_c$, where $T_c$ denotes the deconfinement/confinement
transition temperature since the present model describes 
only the deconfinement phase except for $T=0$ and $q> 0$. 
While we do not discuss on this phase transition here, 

it is shown up to $T=0$, from theoretical viewpoint.

The values of the effective quark mass $\tilde{m}_q$ given by the last 
equations of (\ref{qmass}) 
are shown in the Fig.~\ref{chemical1} by several dots, and we can see they
precisely coincide with $\mu_0(T)$ which are obtained as above.

At low temperature $T<T_c$, we should consider a model in the confinement
phase. In our model, it is realized for $T=0$ and $q>0$. In this case,
we know that
$\tilde{m}_q$ diverges \cite{GY}
 and this means that we can not observe an isolated quark.
In other words, the theory is in the confinement phase. But we always find
a finite $\tilde{m}_q$ for $T>0$ even if $q>0$ in the present model, 
so our model describes the
deconfinement phase so far as $T$ is finite. Then $\mu_0(0)$ is infinite for
$q>0$, but the behavior of $\mu_0(T)$ for $T>0$ is qualitatively similar
to the one of $q=0$. Then, the qualitative property is well described
by $\mu_0(T)$ of $q=0$.
{An important point is that the chemical potential is inactive for
$\mu < \mu_0$, and there is a threshold where it becomes dynamical
and to generate non-zero $n_b\propto \la\Psi^{\dagger}\Psi\ra$.
This result implies that the chemical potential becomes active when
it exceeds the effective quark mass since the gauge theory is in the 
quark deconfinement phase and there is no bound state of quarks.
In order to make clear this point we consider the  relation
with a possible bound state.}

\vspace{.3cm}
As another possibility, we can consider $\mu_0$ as the threshold of some bound state
of quarks. In the present case, it would be the baryon with the mass $M_B$, then we
will have $\mu_0=M_B/N_c$ \cite{Cohen}.
In order to prove this conjecture, we must include
the baryon in the model by for example introducing the D5 brane. 
In this paper, this point is an open problem and we will examine it 
in the near future.

In \cite{Cohen}, the critical value of the isospin chemical potential
is shown to be equal to the half of the meson mass
in the case of quenched approximation. So this point could be
examined by extending the model to non-Abelian DBI action \cite{EGK}.
In a related direction, an interesting analysis in the Higgs branch has been seen
\cite{AEEG}.

\vspace{.5cm}
\section{Phase transition in $dS_4$  }

\vspace{.3cm}
In this case,
the bulk configuration has a horizon at $r=r_0$ which determines
the 4d cosmological constant $\lambda$ as
$\lambda=4r_0^2/R^4$. Then the both the Minkowski BH embeddings are seen, and
we can expect the change of profiles as in the finite temperature case.
Actually,
a kind of phase transition is seen in the case without the chemical potential
but with gauge field condensate \cite{GIN}.
Here we are solving the embedding equations by introducing the chemical 
potential. 

In this case, we should notice the following points. 
(i) For the Minkowski embedding, 
$\tilde{F}_{\rho t}$ should vanish at the end point $\rho=0$.
(ii) As for the BH embeddings, the end point of the brane is at
the horizon $r=r_0$. Near this point, we find 
$e^{\Phi}\propto q(1-r_0/r)^{-3}$ then
$$\tilde{F}_{\rho t}\propto q^{1/2}n_b(1-r_0/r)^{-1/2}$$
which implies $\tilde{F}_{\rho t}=\infty$ at $r=r_0$. In this case, we can not
find any black hole embeddings.
As a result, the black hole embedding solution
can not be obtained for finite $q$ (finite gauge condensate 
$\langle F_{\mu\nu}^2\rangle$) 
and $n_b(\neq 0)$. 

Meanwhile, for $q=0$ and $n_b\neq 0$,
$$e^{\Phi}=1\, , \quad 
\tilde{F}_{\rho t}\propto (1-r_0/r)\to 0$$
in the limit $r=r_0$, and this is independent of $n_b$.
Then, in the case of $q=0$, we find $\tilde{F}_{\rho t}=0$ on the horizon. It shows 
that the black hole embedding is physical in this case, and we actually could
find such solutions.

\vspace{.3cm}
Therefore, we restrict hereafter to the case of $q=0$ for $dS_4$ model in 
order to allow the physical embeddings of $n_b\neq 0$ case as
the black hole embedding, since it is the only possible
embedding in the case of $n_b\neq 0$. 
On the other hand,
as mentioned above, the Minkowski embedding is allowed only for 
$n_b=0$ (phase (A)). But, 
we remind that this does not necessarily mean $A_0=0$. For this embedding,
we can consider a finite chemical potential,
$A_0=\mu$
where $\mu$ is finite but $n_b=0$.
In this case, $\tilde{F}_{\rho t}=0$ at any point, then we find physical
Minkowski embeddings but there is no dynamical role of chemical potential
since it vanishes completely from the action. As in the above case,
the phase (A) is realized for $0<\mu<\mu_0(\lambda)$, and $\mu_0(\lambda)$
represent the critical curve of the phase transition from phase (A) to (B).
And this curve
is obtained according to the method given for the finite temperature case. 

\begin{figure}[htbp]
\vspace{.3cm}
\begin{center}
  \includegraphics[width=6.5cm]{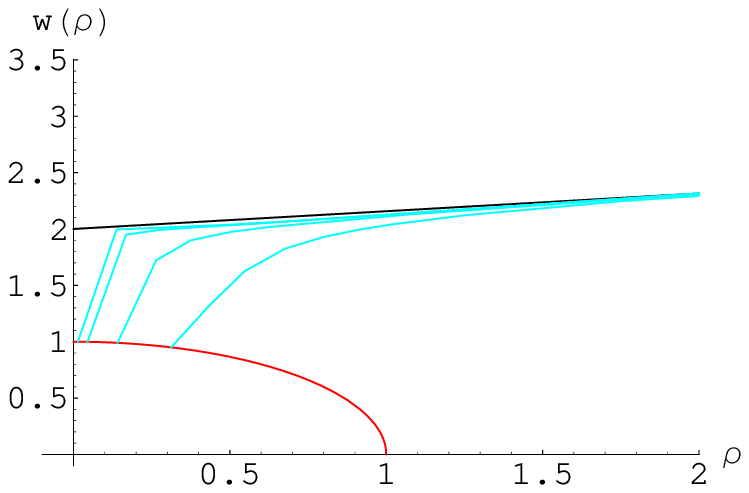}
  \includegraphics[width=6.5cm]{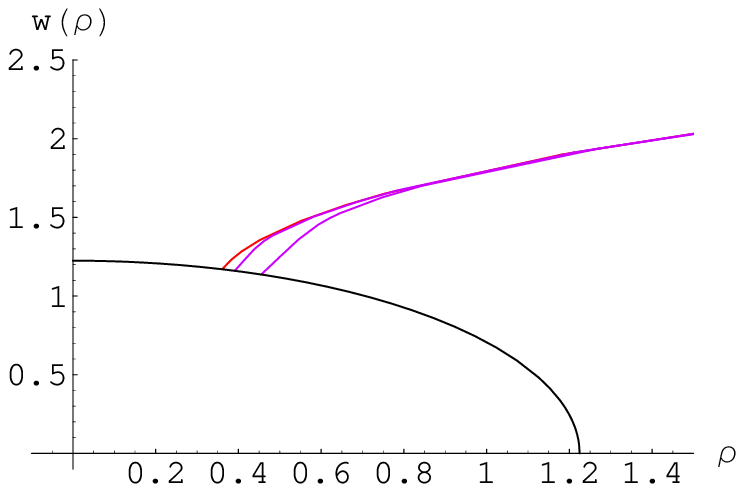}
 \caption{Embedding solutions near the transition point 
for $q=0$, $m_q=2.94966$. The left are for $\lambda=4$, and 
$n_b=$0, 0.000124, 0.00124, 0.0133 and 0.0971 from the above.
The right is for $\lambda=6$, and $n_b=$0, 0.001, 0.01.
 \label{chem4-q0}}
\end{center}
\end{figure}

\vspace{.3cm}
Fixing $m_q$ and $\lambda$,
the typical solutions of $w(\rho)$ for $q=0$ 
near this transition point are shown in the
Fig.~\ref{chem4-q0}.

\begin{figure}[htbp]
\vspace{.3cm}
\begin{center}
  \includegraphics[width=11.5cm]{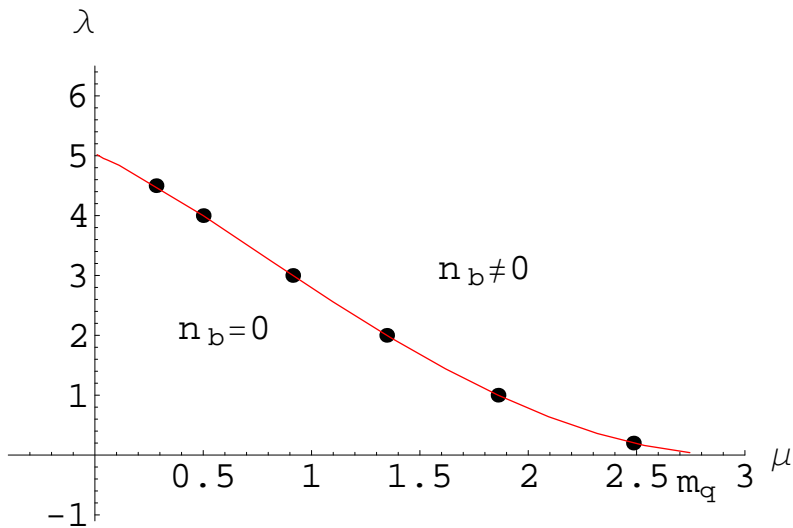}
 \caption{The value of $\mu_0$ for mq=2.94966 and R=1. The dots
 represent the effective quark mass calculated by (\ref{qmass})
 for $m_q=2.94966$.
 \label{mu-lambda}}
\end{center}
\end{figure}

\vspace{.3cm}
In the left of the Fig.~\ref{chem4-q0}, four BH solutions have finite
values of $n_b$ as shown in the figure caption. They have also the corresponding
values of $\mu>\mu_0$, where $\mu_0$ is shown in the Fig.~\ref{mu-lambda}.
For those values of $\mu$, we have also the Minkowski solution given in the 
Fig.~\ref{chem4-q0} with $A_0=\mu$. 
Then, we compare
the D7 energy of each BH embedding solution and the corresponding Minkowski solution
at the same value of $\mu$ in order to see which solution has lower energy.
The D7 energy $E_{\rm D7}$ for $q=0$ which is defined as \cite{GIN}
\beq
S_{\rm D7} =-\tau_7~\int d^7\xi  \sqrt{\epsilon_3}a(t)^3E_{\rm dS_4}
\eeq
then for Minkowski embedding solution,
\beq
E_{\rm dS_4}^M=\int_0^{\infty}d\rho~ \rho^3 A^4
   e^{\Phi}\sqrt{ 1 + (w')^2 }
\ ,
\label{D7-energy}
\eeq
and for BH solution,
\beq
E_{\rm dS_4}^{BH}=\int_{\rho_m}^{\infty}d\rho~ \rho^3 A^4
   e^{\Phi}\sqrt{ (1 + (w')^2)(1-G) }
\ ,
\label{D7-energy-2}
\eeq
where $G$ is given by (\ref{dSG}). 
But as in the case of finite temperature,
$E_{\rm dS_4}$ is also divergent. Although a way to reguralization is 
given in \cite{GIN}, we estimate the difference,
$$ \Delta E\equiv E_{\rm dS_4}^M-E_{\rm dS_4}^{BH}$$
as in the previous section
at the same quark mass $m_q$ and $\lambda$ since $\Delta E$ should be finite.

\vspace{.3cm}
According to this procedure, we find that all the above black hole solutions
give lower energy than the one of the Minkowski embedding, and
the energy decreases with increasing $n_b$. Then the critical
value $\mu_0$ at this $\lambda$ is obtained as the limiting
value of $\mu$ obtained by approaching to $n_b=0$ from BH embedding side.
For $\lambda > $ 5, this limiting value is obtained as zero. Then
the all the embeddings are the BH type even if $n_b=0$ as shown by the right
one of the Fig.~\ref{chem4-q0}.
Performing this procedure for other points of $\lambda <5$, 
we find the critical curve in the $\lambda$-$\mu$ plane. 
The results are shown 
in the Fig.~\ref{mu-lambda}.
As in the high temperature case, we can see the equivalence of $\mu_0$ and 
the effective quark mass $\tilde{m}_q$ obtained from the formula
(\ref{qmass-lam}). Then this result is interpreted as the common property
of the deconfinement phase gauge theory with quarks

\section{Summary}

Here the role of the chemical potential of the quark is studied in terms of the
holographic gauge theory at finite temperature and in $dS_4$. Quarks 
and the chemical potential are introduced
by embedding the probe D7 branes in the bulk configurations  
corresponding to the considering gauge theories. 
In both bulk configurations, there is a horizon. Then two embedding forms, Minkowski 
and black hole, are possible. The chemical potential $\mu$ and the number density $n_b$
are introduced through the Wick rotated
time component of $U(1)$ vector field, $A_0$, on the D7 brane. 

\vspace{.3cm}
In order to embed the D7 brane,
the profile function $w(\rho)$ and $A_0(\rho)$ are solved by the variational principle
of the D7 action. The equations of motion are solved firstly for $w$ by giving $n_b$,
then we obtain $\mu$. This method correspond to study the system in the micro-canonical
ensemble. In this approach, we find a finite $\mu(=\mu_0)$ in the limit of $n_b=0$. 
This implies the existence of 
a parameter region of $0<\mu<\mu_0$ and $n_b=0$, where the D7
brane can not be embedded by the BH embedding. So we need 
Minkowski embedding in this region in order to introduce quarks. 

\vspace{.3cm}
Actually, in
this region the Minkowski embedding is physical since $n_b=0$ and we obtain
the electric potential $A_0=\mu$, then $A_0$ is a constant with respect to
$\rho$. Thus, in this case, we can study the system from the viewpoint of grand canonical ensemble, where $\mu$ is the given parameter. 
So, by varying the chemical potential from zero to large $\mu$, we
find that $\mu$ is inactive below $\mu_0$ and 
$\langle\Psi^{\dagger}\Psi\rangle$ appears for $\mu_0<\mu$ and increases
with $\mu$. 

At the same time with the appearance of $\langle\Psi^{\dagger}\Psi\rangle$,
the profile of the D7 embedding is changed to the BH form.
This is assured by comparing the D7 energies of Minkowski and BH embeddings
with the same $\mu$ and $m_q$ at fixed $T$ (or $\lambda$). The energy of
the BH embedding with finite $n_b$ is always smaller than the one of Minkowski
embedding.
This is therefore a kind of phase transition with the order parameter
$\langle\Psi^{\dagger}\Psi\rangle$. In the bulk, this is seen as the
topology changing of the D7 embedded configuration. We assured this
phase transition in the two different theories 
which are in the quark deconfinement
phase. In both cases,
the critical point $\mu_0$ is identified with the effective quark mass
$\tilde{m}_q$. Thus, this result is consistent with the picture that 
the quark in the deconfinement phase moves freely with this
effective mass. 

\vspace{.3cm}
Then the transition stated above is common to the two deconfinement model.
But we observe some differences between the two models. In both cases, we
introduce a parameter $q$ which is representing the gauge condensate 
$\langle F_{\mu\nu}^2\rangle$. And this parameter is responsible to the 
quark confinement. Actually, for $\lambda=0$ and $T=0$, 
the two theory are equivalent and the theory is in the confinement phase. 
On the other hand, for $T>0$ or $\lambda>0$, both theories
changes to the deconfinement phase. However, when $n_b$ is added, 
while the BH embedding 
is possible in the finite temperature theory for any value of $q>0$,
it is forbidden for dS$_4$.

Another difference is a phase transition which is seen only for the finite
temperature case near very small $\mu$, where a embedding configuration
change is seen between the same BH type configurations. This is not seen
in the case of dS$_4$.

\vspace{.3cm}
The interesting case would be the model in the confining phase, where
the effective quark mass diverges, then $\mu_0$ diverges. 
In other words,
we can not find finite $\langle\Psi^{\dagger}\Psi\rangle$ at any $\mu$. 
The only case,
where $\langle\Psi^{\dagger}\Psi\rangle$ is seen in the confinement phase 
would be in a model in which the baryon is introduced.
In this case, we will find the transition point at
$\mu_0=M_{B}/N_c$ through the same analysis, where $M_B$ is the lowest
baryon mass. On this point we will discuss in the future.

\vspace{.3cm}
\section*{Acknowledgments}
The authors thank to having a chance of discussion with the members of YITP conference 
held at Kinki Univ. Aug. 2007. K. G and M. I thank to S. Nakamura for useful discussions.



\newpage
\end{document}